\documentclass[aps,prb,preprint,showpacs,showkeys]{revtex4-1}   	
\usepackage{graphicx}				
\usepackage{amssymb}
\usepackage{epsfig}

\begin{document}

\title{Superconductivity in a low carrier density system: A single crystal study of cubic Y$_3$Ru$_4$Ge$_{13}$}

\author{Om Prakash, A. Thamizhavel, A.K. Nigam  and S. Ramakrishnan}
\affiliation{Dept. of Condensed Matter Physics and Material Science,\\
Tata Institute of Fundamental research,\\Dr. Homi Bhabha Road, Colaba,, Mumbai 400005, India}


\date{\today}							

\begin{abstract}
We have successfully grown the single crystal of a low carrier density system Y$_3$Ru$_4$Ge$_{13}$ which crystallizes in the cubic crystal structure with the space group {\it{Pm3n}}. Y$_3$Ru$_4$Ge$_{13}$ exhibits super conductivity below 2.85 K as determined from the electrical resistivity, magnetic susceptibility and heat capacity measurements.
The bulk measurements indicate multiband superconductivity in this low carrier system. Hall effect measurements show that the Hall coefficient R$_H$ is positive and it decreases rapidly with temperature which is highly unusual. 
The simple estimation of the carrier concentration gives a value ($~$10$^{20}$/cm$^3$) which is 2-3 orders of magnitude lower than that of a conventional metal like Cu which seems to suggest that Y$_3$Ru$_4$Ge$_{13}$ is a semimetal that display multiband superconductivity. \\
\end{abstract}
\pacs{:~74.70.-b, 74.25.F-, 71.20.Gj
}
\keywords{Supeconductivity, Semimetal, Hall effect.}


\maketitle
\section{Introduction}
A variety of unconventional superconductors present low density of the charge carriers as a common factor, implying that could be the basis for
a unifying picture to understand the superconductivity in such exotic systems. An extremely low density of charge carriers is one of the characteristic features which is shared by
cuprates, fullerenes and MgB$_2$ \cite{r1,r2,r3}. In the case of MgB$_2$, the key role for superconductivity is played by the few carriers of the  $\sigma$  bands. This is quite surprising since low carrier density is an unfavourable element for superconductivity within the conventional framework of BCS or Migdal${-}$Eliashberg theories. Moreover, a small superfluid density, when not in the presence of additional charges not involved in the Cooper pairing (such as the $\pi$-states in the case of MgB$_2$), is unavoidably related to poor screening and to strong electronic correlations, ingredients which are expected to be also detrimental for conventional superconductivity. On these grounds it is hard to understand why these low carrier materials are the best superconductors. As far as
the superconductivity exhibited by inter-metallic compounds are concerned, the role of electron-phonon interaction cannot be overlooked. However, one may have to look beyond the
conventional framework of BCS or Migdal${-}$Eliashberg theories in order to understand the unconventional superconductivity in these compounds. From the experimental side, it is important to look for new superconducting materials with low carrier density. Recently, 
unconventional superconductivity has reported in a few compounds such as Rh$_{17}$S$_{15}$ \cite{r4} and YPtBi \cite{r5}, the latter one has also a noncentrosymmetric crystal structure. 
A series of intermetallic compounds with no metalloids was reported by Remeika {\it et al} and Hodeau {\it et al} \cite{r6}. These compounds (R$_3$Rh$_4$Sn$_{13}$) crystallize in cubic structure
{\it{Pm3n}}. Amongst these compounds, those of La, Yb, and Th are superconducting whereas for Gd and Eu compounds a magnetic transition was observed. Efforts were made to look for similar systems with germanides which can display superconductivity and magnetism. Segre and Braun \cite{r7} were successful in reporting superconductivity and magnetic order in R$_3$Ru$_4$Ge$_{13}$ which crystallizes in the same {\it{Pm3n}} structure. Our studies \cite{r8,r9,r10} on polycrystalline alloys
have shown that R$_3$Ru$_4$Ge$_{13}$  (R=Y, Ce, Pr, Nd, Ho, Er, Dy, Yb, Lu) series  exhibit  unusual physical properties which could be considered as those belonging to low band gap semiconductors. We have also shown that \cite{r10} the compounds having magnetic rare earth element show paramagnetic behaviour down to  1.5 K whereas,  Lu$_3$Ru$_4$Ge$_{13}$ and Y$_3$Ru$_4$Ge$_{13}$ exhibit Pauli paramagnetism above 4.2 K. They also undergo a superconducting transition at 2.3 K and 1.8 K, respectively. The resistivity of all the samples except that of Yb$_3$Ru$_4$Ge$_{13}$ increases with the decrease of temperature from 300 K down to 
1.5~K.  Later  investigation  showed that these compounds can also be classified as good thermoelectric materials \cite{r11}. There is also another reason to look for new low carrier density semimetals that may exhibit
superconductivity. Earlier theoretical investigation \cite{r12,r13} demonstrated that in a donor-doped multivalley semiconductor or multivalley semi-metal, at very high magnetic fields, there exists a critical temperature below which a
new triplet superconducting phase must arise. This phase is  a mirror image of the spin-up, spin-down  Cooper-pair condensation  where the spins are replaced by the  indices of the valleys. 
Because the magnetic field does not couple to the indices this transition is strongly enhanced in the presence of a magnetic field. There has been some attempts to observe this effect in a semimetal
like Bismuth \cite{r14}. It is worthwhile to look such exotic superconducting state in other semimetals having lower carrier density.
\section{experimental details}
\noindent
Single crystals of Y$_3$Ru$_4$Ge$_{13}$ were grown using Czochralski crystal pulling method in a tetra-arc furnace under high purity Argon atmosphere. The starting materials were of high purity Y(99.9\%), Ru(99.999\%) and Ge(99.999\%). A total of 10 grams of the stoichiometric Y$_3$Ru$_4$Ge$_{13}$ was taken and melted several times in the tetra-arc furnace to make a homogeneous polycrystalline mixture. Crystal was pulled using a Tungsten seed rod at the rate of 10 mm/hour for a total duration of 5-6 hour to get 5~cm long crystal. The diameter of the crystal was 3-4~mm. The phase purity was characterised by powder X-ray diffraction using PANanalytical x-ray diffractometer utilising monochromatic Cu-K$\alpha$ radiation with the wavelength 1.5406~\AA. The EPMA and EDAX were performed on polished surfaces and show single phase and correct stoichiometry of ($3$-$4$-$13$). The single crystal was oriented along the crystallographic direction [110] by means of the Laue back reflection using a Huber Laue diffractometer and cut to desired dimensions using a spark erosion cutting machine. Y$_3$Ru$_4$Ge$_{13}$ has a cubic {\it{Pm3n}} structure with 40 atoms per unit cell (2 formula unit). It also contains two inequivalent Ge sites. One can visualize  Y$_3$Ru$_4$Ge$_{13}$ structure as a unit consists of three substructures: edge-sharing Ge1(Ge2)$_{12}$ icosahedra, Y-centered cubotahedra R(Ge2)$_{12}$, and corner-sharing  Ru(Ge2)$_6$ trigonal prisms. The corner-sharing Ru(Ge2)$_6$  trigonal prisms create "cages" containing a Ge1 atom similar to the cages observed in Skutterudites \cite{r15}.
The unit cell of the cubic structure Y$_3$Ru$_4$Ge$_{13}$, space group {\it{Pm3n}} is shown in Fig. 1.
\begin{figure}[htbp]
\begin{center}
\epsfig{figure=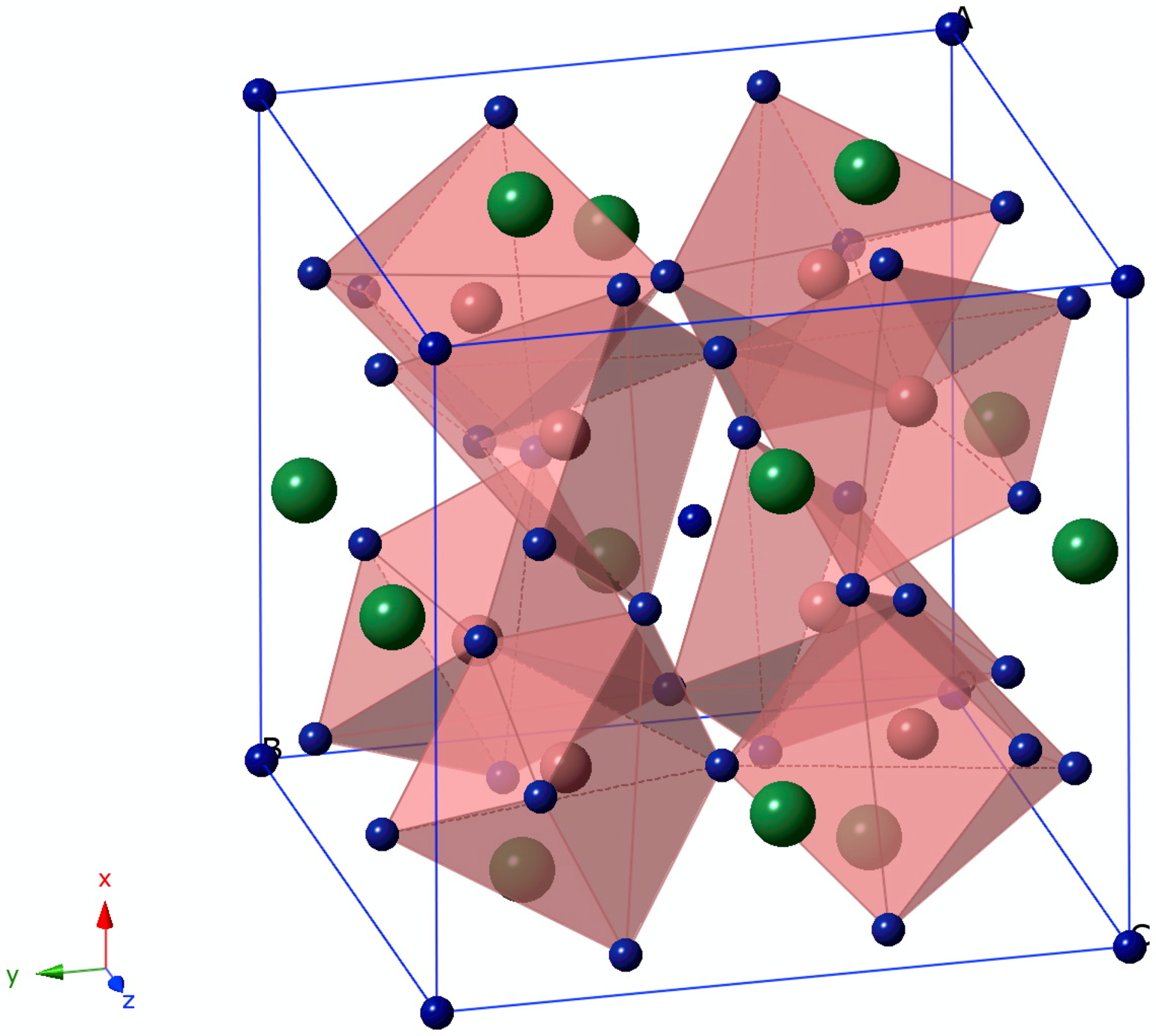,scale=0.6,angle=-90}
\caption{(Color online) Crystal structure of Y$_3$Ru$_4$Ge$_{13}$. Yttrium atoms are shown in green in 6d position, Ruthenium are shown in light pink in 8e position and Germanium in 2a and 24k positions are shown in
dark blue which are two inequivalent Wyckoff positions.}
\label{fig1}
\end{center}
\end{figure}
The Rietveld fit \cite{r16} to the powder X-ray diffraction data is shown in Fig. 2.
\begin{figure}[htbp]
\begin{center}
\epsfig{figure=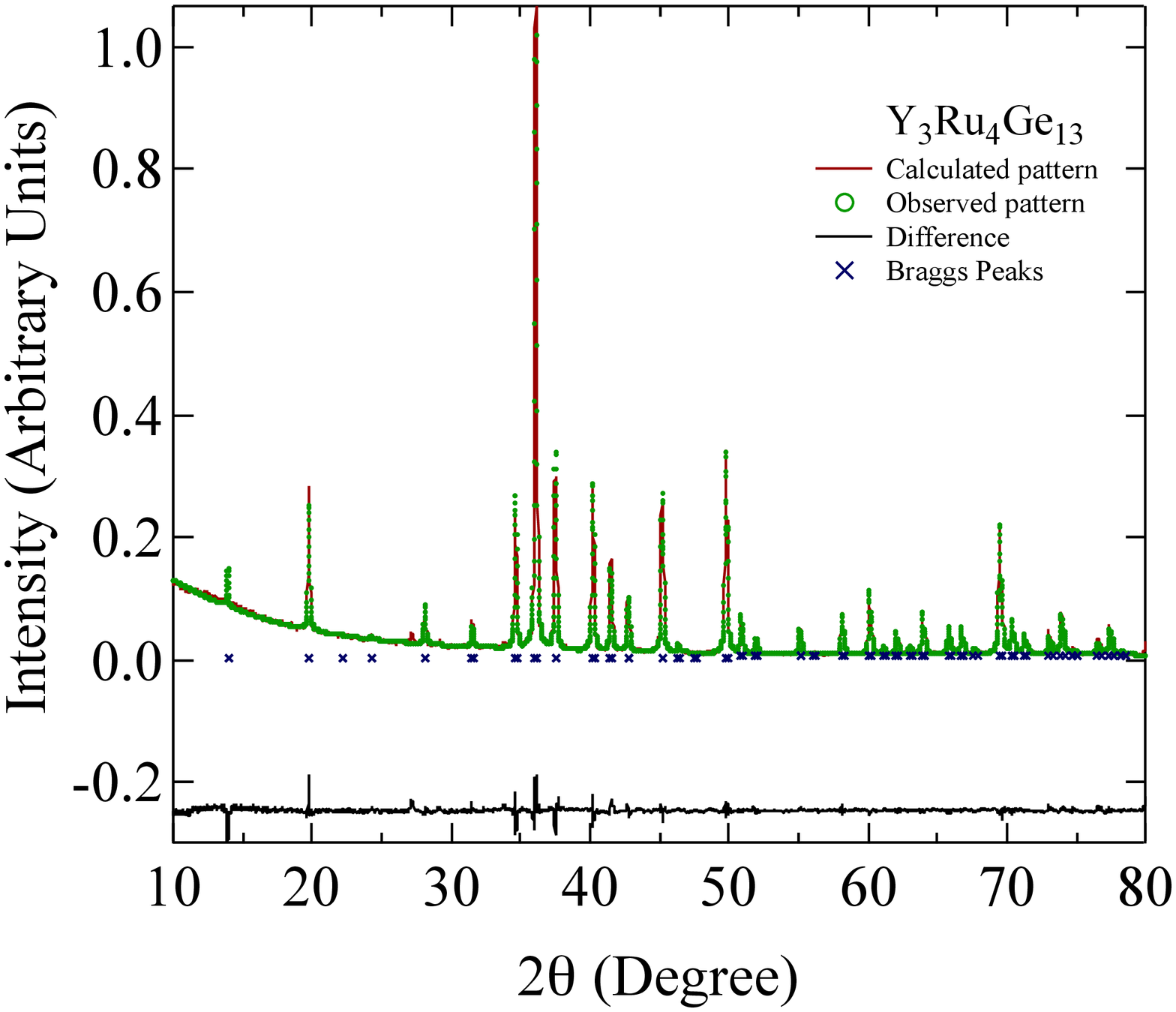,scale=0.6,angle=90}
\caption{(Color online) Powder x-ray diffraction data of the powdered single crystal of cubic Y$_3$Ru$_4$Ge$_{13}$. The solid black line is the simulated data using FullProf (Rietveld program).}
\label{fig2}
\end{center}
\end{figure}
The values for the lattice constants estimated from the fit are  {\it{a}}~=~{\it{b}}~=~{\it{c}}~=~8.9723($\pm$0.0004)\AA. These values are in good agreement with the values reported earlier experimental and theoretical investigations \cite{r10,r17}.
A commercial SQUID magnetometer (MPMS5, Quantum Design, USA) was used to measure the temperature dependence of the magnetic susceptibility ($\chi$) from 1.8 to 300~K. The resistivity was measured using a four-probe ac resistance measurement technique using a home-built setup and the contacts were made using silver paint on a bar shaped 0.5 mm thick, 3.9 mm long and 2.7 mm wide sample. The temperature was measured using a calibrated Si diode (Lake Shore Inc., USA) temperature sensor. The sample resistance was measured with and LR370 AC Resistance Bridge (Lake Shore Inc. USA). The absolute resistivity has an error of 2\% due to errors in the estimation of the geometrical factors of the sample. The heat capacity was measured (with an accuracy of 1\%) using a commercial setup (PPMS, Quantum Design, USA) in the temperature range from 1.9 to 100~K in zero magnetic field as well as in a field of 1 T field. AC-$\chi$ measurement was done in a home-built set up using a very sensitive mutual inductance bridge (sensitivity of  10$^{-7}$ emu/gm).
\section{Results and Discussion}
\subsection{\label{A.}Resistivity Studies}
Figure (3) shows the temperature dependence of the resistivity  of Y$_3$Ru$_4$Ge$_{13}$ crystal  along the [110] direction from 1.6 to 300~K. The inset (a) shows the  low temperature data 
displaying the  superconducting transition  at 2.85~K. 
The  normal state resistivity of the Y$_3$Ru$_4$Ge$_{13}$ decreases with increasing temperature from 3 to 300~K indicating semimetallic nature of the sample. This is in  broad agreement with the 
previous report for the polycrystalline sample of Y$_3$Ru$_4$Ge$_{13}$ \cite{r10}. However, present single crystal has a resistivity (~90 $\mu$$\Omega$ cm at 3~K) which is an order of magnitude smaller than that of the polycrystalline sample (~1.1m$\Omega$ cm at 3~K). We believe that the latter had significant disorder which lead us to erroneously conclude \cite{r10} that Y$_3$Ru$_4$Ge$_{13}$ is a low band-gap semiconductor. The present resistivity
data in conjunction with other  bulk measurements suggest that the sample is a semimetal.
\begin{figure}[htbp]
\begin{center}
\epsfig{figure=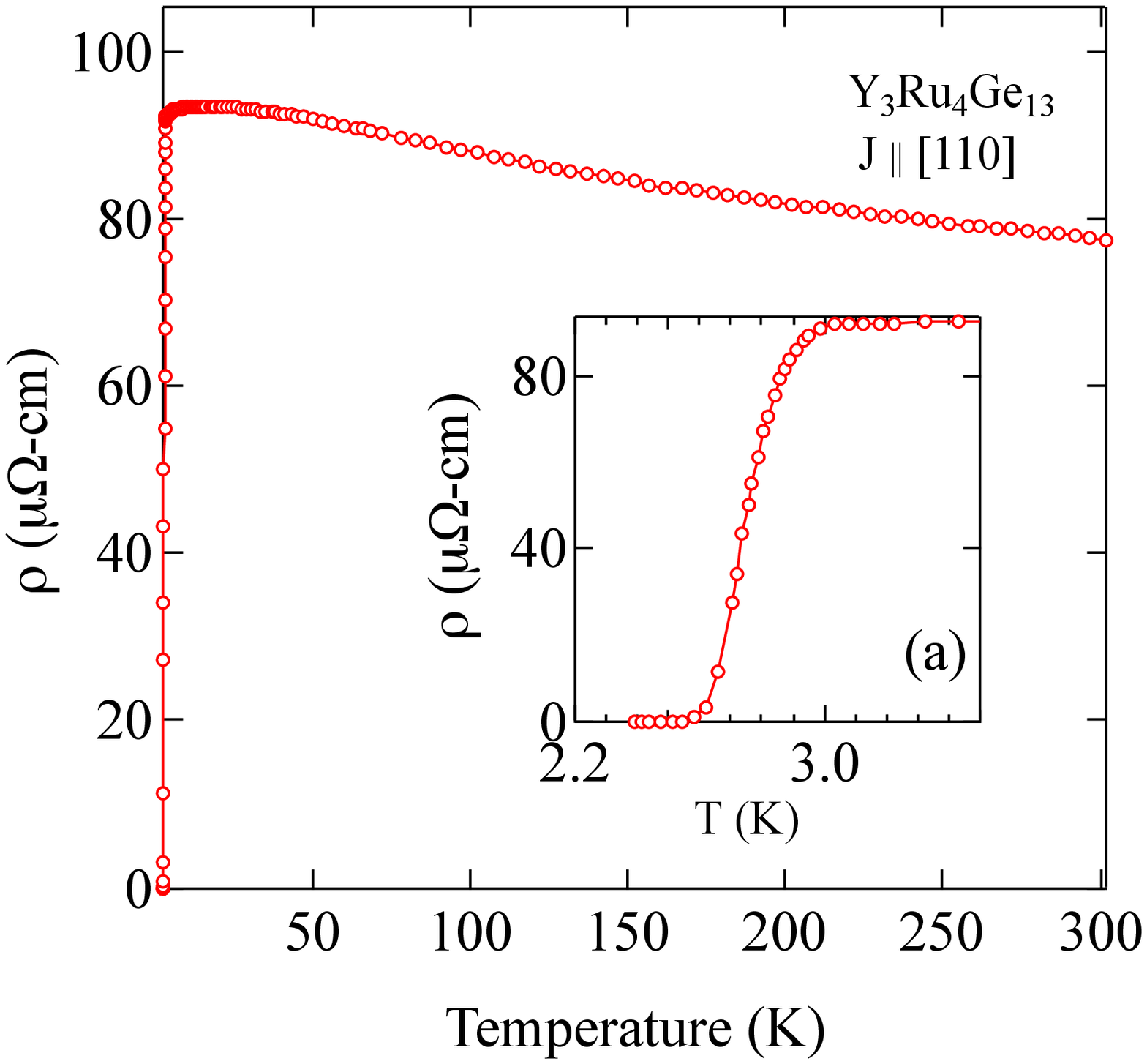,scale=0.7}
\caption{(Color online) Temperature dependence of the electrical resistivity ($\rho$) along the (110) direction of cubic (Pm3n) Y$3$Ru$4$Ge$13$. Inset (a) shows the superconducting transition at 2.85 K.}
\label{fig3}
\end{center}
\end{figure}
\subsection{\label{B.}Magnetic Susceptibility Studies}
The temperature dependence of the magnetic susceptibility $\chi(T)$ of Y$_3$Ru$_4$Ge$_{13}$ measured in a field of 1~T is shown in Fig. 4.
The inset (a) shows the low temperature susceptibility (along the [110] direction) data at 2~mT field displaying  the superconducting (diamagnetic) transition at 2.85~K, which is in agreement with the resistivity data. The zero-field-cooled (ZFC) and field-cooled (FC) susceptibility data indicate significant amount of pinning of the vortices in the sample. The inset (b) depicts the AC susceptibility data which signifies the diamagnetic screening due to the superconducting transition at 2.85~K. The volume susceptibility is close to the value of $-1$ showing nearly perfect diamagnetism (Meissner effect) in the superconducting state. Y$_3$Ru$_4$Ge$_{13}$ exhibits Pauli paramagnetism in the normal state. The value of $\chi$ increases from (3$\pm$0.1)$\times$10$^{-4}$ emu/mol  at 300~K to (16.8$\pm$0.1)$\times$10$^{-4}$ emu/mol at 3~K before it becomes a superconductor. 
 \begin{figure}[htbp]
\begin{center}
\epsfig{figure=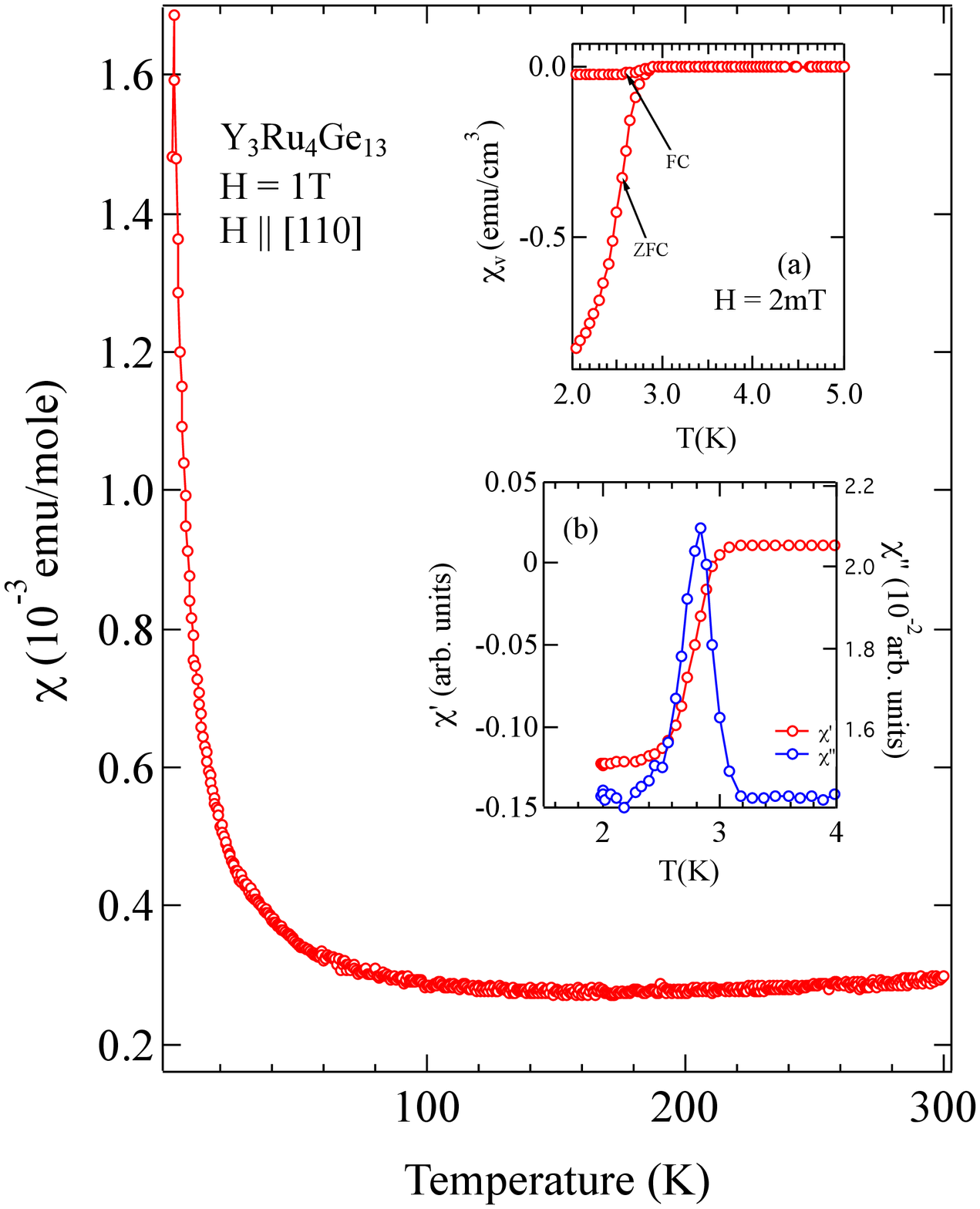,scale=0.5,angle=0}
\caption{(Color online) Temperature dependence of the magnetic susceptibility ($\chi$) along the [110] direction of cubic {\it{(Pm3n)}} Y$_3$Ru$_4$Ge$_{13}$. Inset (a) shows superconducting transition at 2.85~K (FC and ZFC data). Real and Imaginary components of Ac-Susceptibility are shown in inset (b).}
\label{fig4}
\end{center}
\end{figure}
In general, the observed value of the susceptibility can be written as,
\begin{equation}
\chi_{obs}~=~\chi_{core}~+~\chi_{Landau}~+~\chi_{Pauli}
\end{equation}
$\chi_{core}$ is the core diamagnetism, $\chi_{Landau}$ is the Landau diamagnetism and  $\chi_{Pauli}$ is the Pauli paramagnetism. The equation (1) can be rewritten as,
\begin{equation}
 \chi_{obs}~-~\chi_{core}~=~\chi_{Pauli} \left[1-\frac{1}{3}\left(\frac{m}{m_{b}}\right)\right]
\end{equation}
Here,  $\chi_{Pauli}$ = N$_A$~$\mu_B^2$~S~N(E$_F$) where N$_A$ is the Avogadro number, $\mu_B$ is the Bohr magneton, S is the Stoner factor, N(E$_F$) is the density of states at the
Fermi level, m is the free electron mass and m$_b$ is the band mass. The estimated value of $\chi_{Pauli}$ is 9.68$\times$10$^{-4}$ emu/mol. If one assumes the core diamagnetic susceptibility of Y as -24$\times$10$^{-6}$ emu/mol , Ru as -18$\times$10$^{-6}$ emu/mol and Ge as -9.22$\times$10$^{-6}$ emu/mol, we get a value -~3.56 $\times$10$^{-4}$ emu/mol as the total contribution to the core diamagnetism in Y$_3$Ru$_4$Ge$_{13}$. The value of $\chi_{obs}$ is 2.97$\times$10$^{-4}$ emu/mol at 300 K. From this we calculate $\chi_{Pauli}$ to be 9.69$\times$10$^{-4}$ emu/mol. In this calculation the value of the ratio m/$m_b$ is taken as 1.
From this we get a value of the Stoner enhancement factor S as 1, which confirms the sample is non magnetic. So for a nonmagnetic sample like Y$_3$Ru$_4$Ge$_{13}$ , one expects a temperature independent susceptibility.
Hence, the small increase in $\chi$(T)  could be attributed the presence of paramgetic impurities ($<$ 50~ppm in Y) in the starting material used in the preparation of the single crystal of Y$_3$Ru$_4$Ge$_{13}$.
\subsection{\label{C.}Heat Capacity Studies}
The temperature dependence of the heat-capacity ($C_p$) from 0.4 to 4~K of Y$_3$Ru$_4$Ge$_{13}$ is shown in Fig. 5 at 0~T, 1~T and 2~T magnetic fields respectively.
The jump in $C_p$ in the absence of magnetic field (0~T) at 2.85~K ($\Delta$C$\approx$17 mJ/mol-K) signifies bulk superconductivity in the sample below 2.8~K. The temperature dependence of 
the $C_p$ is fitted to the expression
\begin{equation}
\frac{C_p}{T}~=~\gamma~+~\beta~T^2~
\end{equation}
where $\gamma$ is due to the electronic contribution and $\beta$ is due to the lattice contribution to the heat capacity.
\begin{figure}[htbp]
\begin{center}
\epsfig{figure=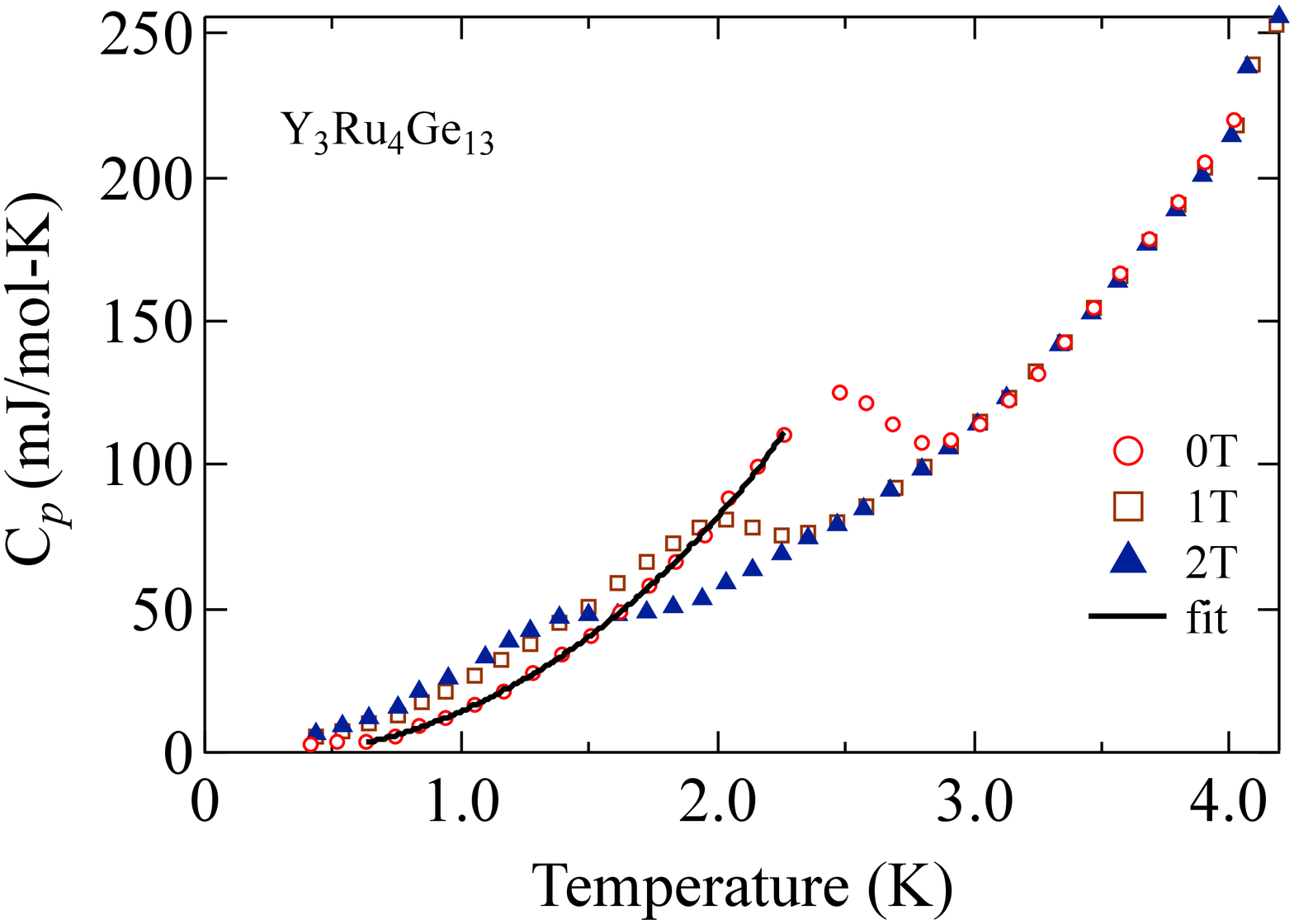,scale=0.5,angle=90}
\caption{(Color online) Temperature dependence of the heat-capacity $C_{p}$  of Y$_3$Ru$_4$Ge$_{13}$  in a magnetic field of 0~T, 1~T and 2~T from 0.4 to 4~K. The zero field data is fitted to the equation (5) from 0.6 to 2.4~K.}
\label{fig5}
\end{center}
\end{figure}
\begin{figure}[htbp]
\begin{center}
\epsfig{figure=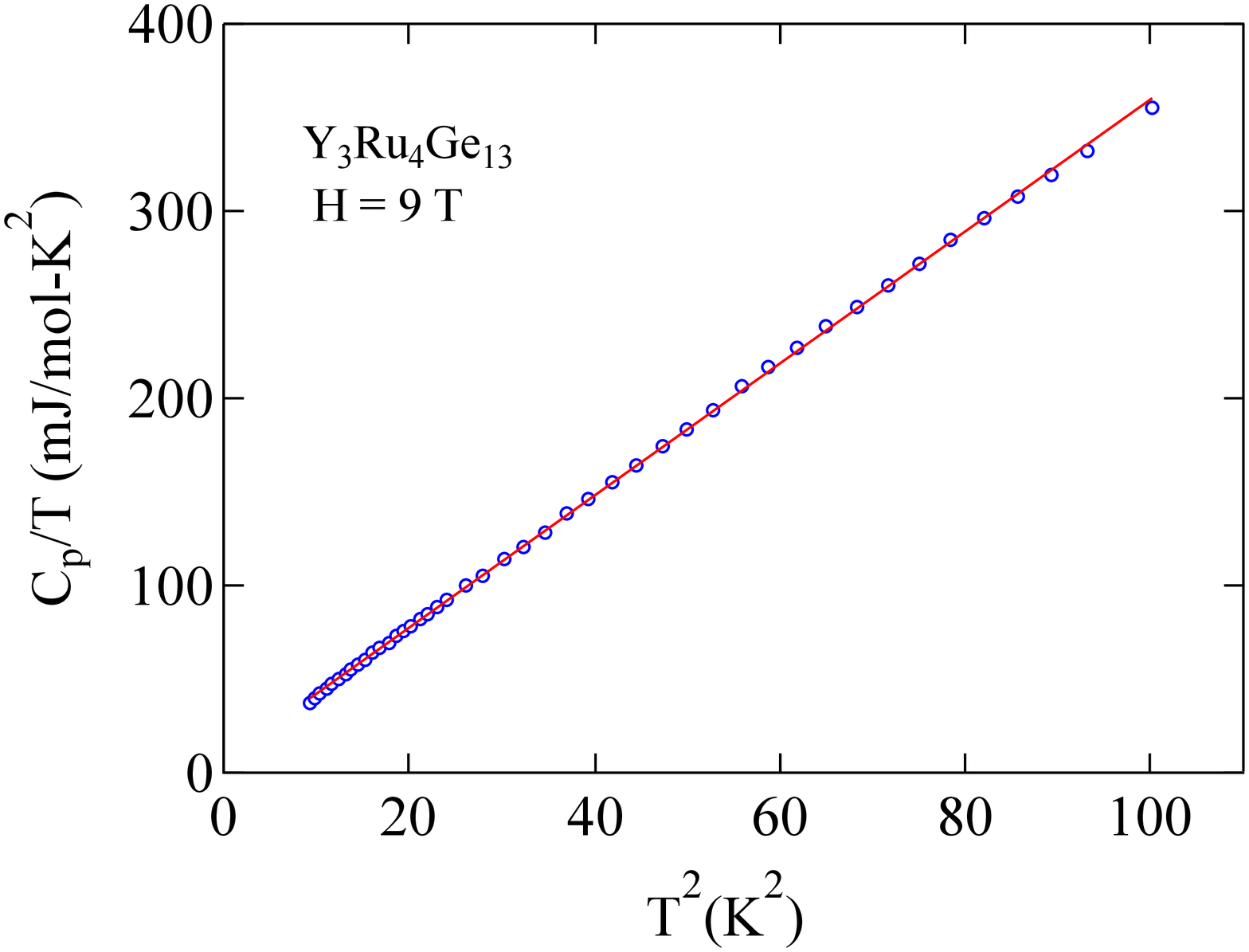,scale=0.5,angle=90}
\caption{(Color online) Plot of $C_p$/T vs T$^2$ at 9~T. The solid line is the fitted data to the eq. (3).}
\label{fig6}
\end{center}
\end{figure}
The superconductivity is suppressed by a magnetic field of 9 T and the data is fitted to the eq.~(3), which is displayed as a solid line in Fig.~6. The fit to the heat capacity data using eq.~(3) in the temperature range 2 to 10~K yielded (7.08$\pm$0.38)~mJ/mol-K$^2$ and (3.52$\pm$0.008) mJ/mol-K$^4$ for $\gamma$ and $\beta$ respectively. The value of the ratio ($\Delta$Cp/$\gamma$Tc) is 0.85. The Debye temperature $\theta_D$ can be calculated using the value of $\beta$ (3.52$\pm$0.008) mJ/mol-K$^4$ from the relation,
\begin{equation}
\theta_D=\left(\frac{12\pi^4N_Ank_B}{5\beta}\right)^{1/3}
\end{equation}
where $N_A$ is the Avogadro's number, n is the number of atoms per formula unit, and $k_B$ is the Boltzmann's constant. The value of Debye temperature {$\theta_D$} as calculated using eq.~(4) is 222.9~K. 
The zero field heat capacity data at temperatures below ${T_c}$ is fitted to the equation,
\begin{equation}
C_s=A~exp{(-{\Delta}/k_{B}T)}+C~T^3
\end{equation}
where $\Delta$ is the energy gap related to the Cooper pair binding energy and $k_B$ is the Botzmann's constant. The cubic term is added to fit any phonon contribution to the heat capacity in the superconducting state of the sample. The value of $\Delta$ comes out to be 0.21~meV. The value of 2$\Delta$/{$k_B$$T_{c}$} is 1.7, which is smaller than the value predicted by BCS theory.
\subsection{\label{D.}Hall effect studies }
Hall effect measurements were performed to estimate the carrier density in Y$_3$Ru$_4$Ge$_{13}$. The resistance measurements were performed along the [110] direction of the sample as a function of temperature at different applied magnetic fields. In order to eliminate the longitudinal resistivity contribution to the measured R$_H$ due  to the misalignment of the hall voltage contacts the Hall
resistivity was derived from the antisymmetric part of the transverse resistivity under magnetic field reversal at given temperature, i.e., $\rho_H$~=~[$\rho_H$(+H)~-~$\rho_H$(-H)]/2. From this
Hall coefficient R$_H$~=~$\rho_H$/H has been computed. Figure~(7) shows the Temperature dependence of the Hall coefficient and the estimated value of the carrier density at 9~T from 1.9 to 300~K. 
The electron number density is calculated using the following equation assuming only one type of charge carrier is present in the sample. 
\begin{equation}
R_H=-\frac{1}{ne}
\end{equation}
where R$_H$ is hall coefficient (Coulomb-cm$^3$), e is the charge of electron, and n is the number density of electrons~(/cm$^3$). 
The inset in the same figure
shows unusual filed dependence of R$_H$ which gets more prominent at higher temperatures. In the case of Y$_3$Ru$_4$Ge$_{13}$, it is possible that the temperature
of the carrier concentration or mobility can show the increase in the dependence of R$_H$ with H at higher temperatures. However, the robustness of the value
of the field (20~mT) where R$_H$ displays a maximum is independent of the temperature. This low field behaviour of R$_H$ is not understood at this moment. 
The magnetic field dependence of the Hall coefficient at different temperature in low fields is shown as an inset (a) in Fig.~7.
\begin{figure}[htbp]
\begin{center}
\epsfig{figure=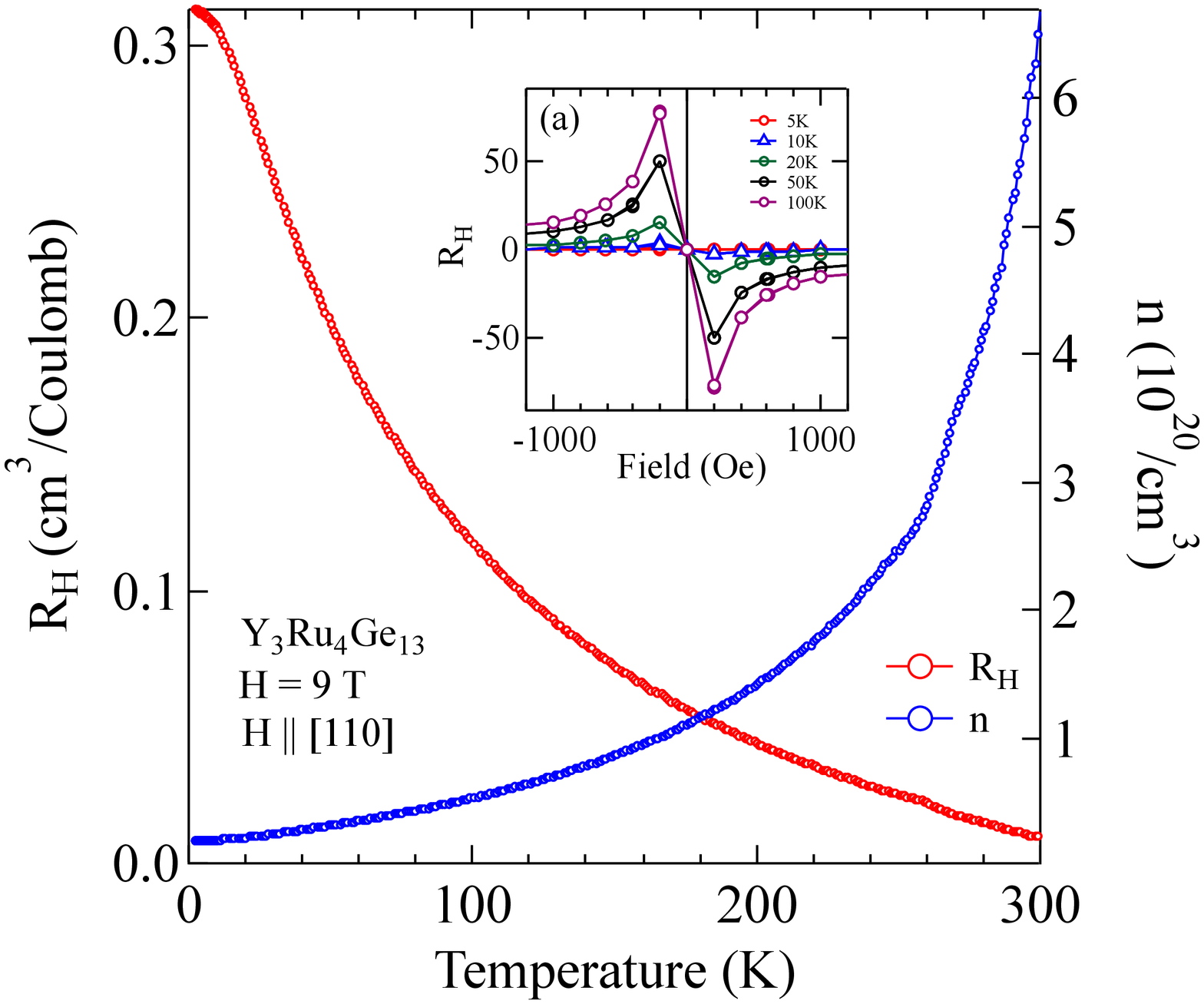,scale=0.5,angle=90}
\caption{(Color online) Temperature dependence of the Hall coefficient and carrier concentration in Y$_3$Ru$_4$Ge$_{13}$.}
\label{fig7}
\end{center}
\end{figure}
We see a significant rise in the carrier concentration at higher temperatures. The observation of the temperature dependence of the Hall coefficient is not unusual in multiband
semimetal like Y$_3$Ru$_4$Ge$_{13}$ which has both electrons and holes.  Significant temperature dependence of the Hall coefficient could arise either
due to the different temperature dependence of the mobility of the charge carriers in each band or due to the thermal expansion of the material which leads to
a change in the  distribution of charge carriers amongst the bands. Even though strong temperature dependence of R$_H$ is theoretically possible, such behaviour has been
confined to few systems. For example Hall coefficients of both High-T$_c$ cuprates and layered compound like MgB$_2$ display inverse temperature dependence. In the case
of Y$_3$Ru$_4$Ge$_{13}$ crystal, one sees an unusual exponential rise in R$_H$ at low temperatures  before it becomes a superconductor. Clearly the electronic
structure of Y$_3$Ru$_4$Ge$_{13}$ is quite complex as shown by the preliminary band structure calculations \cite{r17}.
\subsection{\label{E.}Upper Critical Field studies }
The upper critical field is estimated by measuring the resistance of the sample under constant magnetic field. The superconducting transition temperature is defined as the temperature corresponding to the midpoint of the resistance jump. The temperature dependence of the critical magnetic field $Hc_2$ along the [110] direction of the sample is shown in Fig.10. In non-magnetic superconductors the magnetic field interacts with the conduction electrons basically through two different mechanisms. Both lead to pair breaking and eventually destroying the superconductivity at critical magnetic field. One of these arises due to the interaction of the magnetic field with the orbital motion of the electrons leading to what is known as orbital pair breaking and the other is due to the interaction of the magnetic field with the electron spins (Pauli paramagnetic limiting effects). Orbital pair breaking is the dominant mechanism at low fields and Pauli paramagnetic effect limits the critical field at very high magnetic fields. We have fitted the temperature dependence of $Hc_2$ determined by the orbital and spin-paramagnetic effect to the theory of dirty type-II superconductor, using Werthamer-Helfand-Hohenberg formula\cite{r18},
\begin{eqnarray}
ln\left(\frac{1}{t}\right)=\sum_{\nu=-\infty}^{\nu=\infty}\Biggl\{\frac{1}{|2\nu+1|}-\Biggl[{|2\nu+1|}+
\frac{\overline h}{t}+ \frac{(\alpha\overline h/t)^2}{|2\nu+1|+(\lambda_{so}+\overline h)/{t}}\Biggr]^{-1}\Biggr\}
\end{eqnarray}
where
\begin{eqnarray}
t=\frac{T}{T_c}, ~~    
\overline h = \frac{4H_{c_2}(T)}{\pi^2T_c \bigg| \frac{dH_{c_2}(T)}{dT}\bigg|_{T_c}}
\end{eqnarray} 
 $\alpha$ is Maki parameter, and $\lambda_{so}$ is the spin-orbit scattering constant. We obtain a value of (2.06$\pm$0.14) for $\lambda_{so}$ from the fit.
The Maki parameter is calculated using the formula\cite{r17}
\begin{equation}
\alpha=-0.528\bigg| \frac{dH_{c_2}(T)}{dT}\bigg|_{T_c}
\end{equation}
We get a value of 1.32 for $\alpha$ from the eq.(9).
The value of ${\left|{dHc_2(T)/dT}\right|_{T_c}}$ as obtained from the slope of $Hc_2$ vs T curve at $T_c $ is $2.49\pm 0.07$. The upper critical field at zero temperature $H_{c_2}$(T=0) can be estimated from our data using the WHH approximation\cite{r18},
\begin{equation}
H_{c_2}(0)=0.69T_c\bigg| \frac{dH_{c_2}(T)}{dT}\bigg|_{T_c}
\end{equation}
From this equation we get the value of $4.92\pm 0.34$ T for $Hc_2(0)$ which is very close to the extrapolated  experimental value of 4.83 T. However, assuming the dirty limit for a type-ll superconductor, one can also estimate the ${dH_{c_2}(T)/dT}$ using the relation\cite{r19},
\begin{equation}
\frac{dH_{c_2}}{dT} =  4.48*10^4{\gamma}{\rho_{\Omega cm}} (Oe K^{-1})
\end{equation}
where {$\gamma$} is the electronic heat capacity coefficient (ergs/cm$^3$-K$^2$), {$\rho_{\Omega cm}$} is the residual resistivity. Using this formula we get ${dH_{c_2}/dT}$=1358 Oe/K, which is much smaller than the value(2.5 T/K) obtained from the experiment.
\begin{figure}[htbp]
\begin{center}
\epsfig{figure=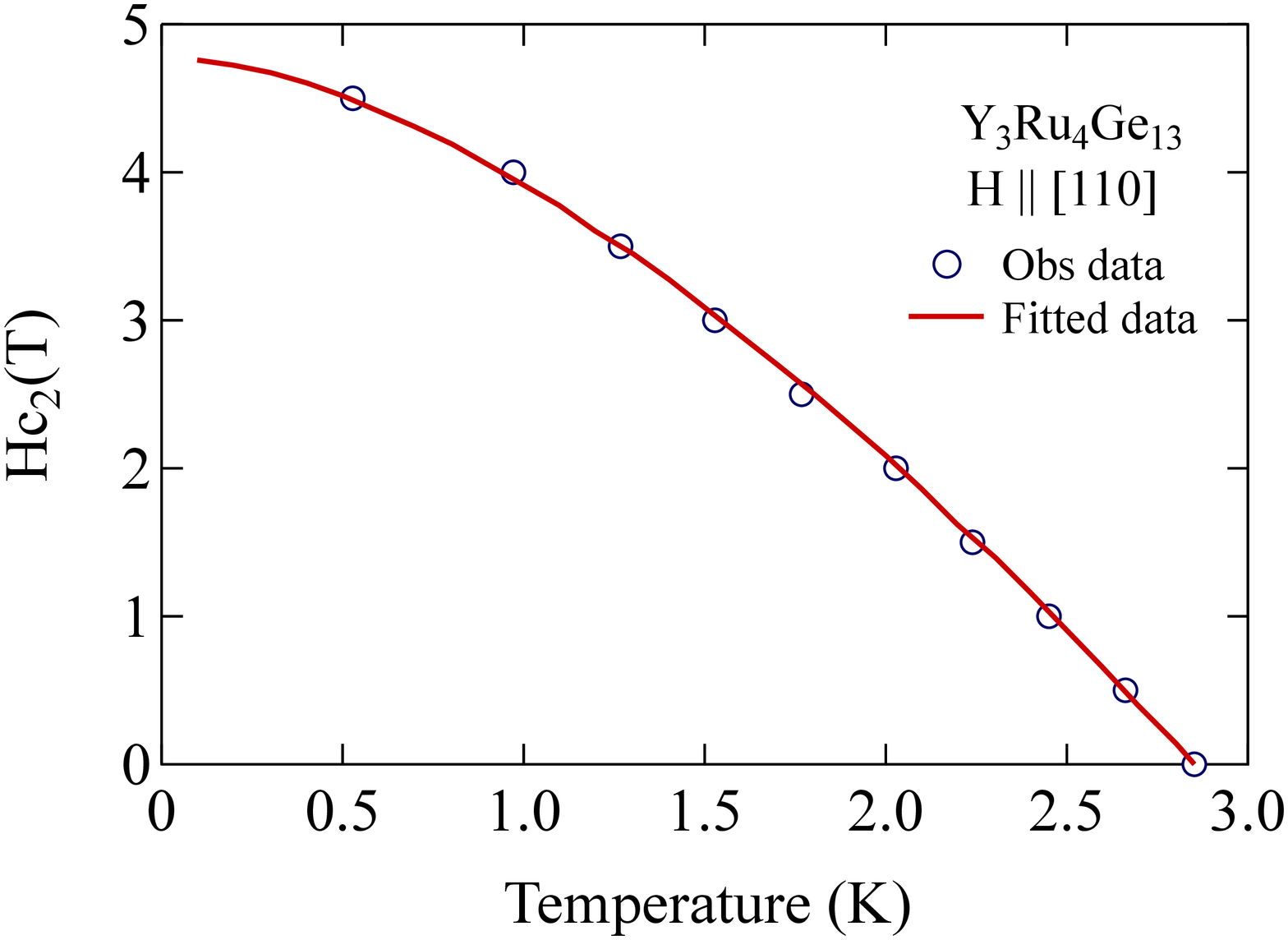,scale=0.5,angle=90}
\caption{(Color online) Temperature dependence of the upper critical field ($H_{c_2}$) of Y$_3$Ru$_4$Ge$_{13}$ along the [110] direction. The data is fitted using WHH equation for dirty superconductors.}
\label{fig8}
\end{center}
\end{figure}
\section{Estimation of normal and superconducting state parameters}
Using the values of {$\theta_D$} and $T_c$, we can estimate the electron-phonon scattering parameter,
 {$\lambda$}, from McMillan's theory, where {$\lambda$} is given by\cite{r20}
\begin{equation}
\lambda=\frac{1.04+\mu^*ln(\theta_D/1.45T_c)}{(1-0.62\mu^*)ln(\theta_D/1.45T_c)-1.04} 
\end{equation}
Assuming  $\mu^*$=0.13, we find the value of $\lambda$ to be 0.59. We can calculate the thermodynamic critical field $H_c(0)$ from the expression\cite{r19},
 \begin{equation}
H_c(0)=4.23\gamma^{1/2}T_c
\end{equation}
where ${\gamma}$ is the heat capacity coefficient (mJ/mol-K$^2$). This gives a value of $H_c(0)$ as 218~Oe which is in agreement with the experiments. We can also estimate the Ginzburg-Landau coherence length {$\xi_{GL}$} at T~=~0~K from the relation,
\begin{equation}
\xi_{GL}(T) = \left(\frac{\Phi_0}{2\pi H_{c_2}(T)}\right)^{1/2}
\end{equation}
We got a value of 393 {\AA} for {$\xi_{GL}$}(0).
We can also calculate $\kappa$(0) in the dirty limit using the relation, 
\begin{equation}
\kappa_{GL}(0) = {H_{c_2}(0)\over \sqrt{2}~H_c(0)}
\end{equation}
The value of {$\kappa_{GL}$}(0)  come out to be 12.6. The Ginzburg-Landau penetration depth {$\lambda_{GL}$}(0) for the dirty superconductors at 0K can be calculated using the relation,
\begin{equation}
\lambda_{GL}(0) =\xi_{GL}(0)\kappa_{GL}(0)
\end{equation}
which gives the value of {$\lambda_{GL}$}(0) = 4951~{\AA}. The lower critical field value can be calculated using the relation,
\begin{equation}
H_{c1}(0)=\frac{H_c(0)ln[\kappa(0)] }{2^{1/2}~\kappa(0)}
\end{equation}
which gives a value of 31~Oe for the lower critical field at 0~K. This value of $H_{c1}$ is in agreement with the value obtained from the magnetisation measurements.
The enhanced density of states can be calculated using the relation,
\begin{equation}
N^*(E_F) = 0.2121 \gamma /N 
\end{equation}
where N is the number of atoms per formula unit and $\gamma$ is expressed in mJ/mol K$^2$. The value of $N^*(E_F)$ is 0.075 states /(eV atom spin direction) and the value of bare density of states N(E$_F$)~=~N$^*$(E$_F$)/(1+$\lambda$)= 0.047 states /(eV~atom~spin-direction). 
\begin{table*}[htdp]
\caption{Normal and superconducting state properties of a single crystal Y$_3$Ru$_4$Ge$_{13}$}
\begin{ruledtabular}
\begin{tabular}{ccccccccccccccc}
${\rho_{300K}}$ & $\rho _{4.2K}$ & $\gamma$  & $\theta _D$  & $H_{c_2}$(0) & $H_{c1}$(0) & $dH_{c_2}$/dT & $\kappa_{GL}(0)$ & $\lambda$ & $\xi_{GL}(0)$ & $\lambda_{GL}$(0)  \\
$\mu \Omega$-cm &  $\mu \Omega$-cm & ~~mJ/mol-K$^{2}~$ &  K~  & T & T & T/K & & & nm & nm  \\
\hline 
77  & 93  & 7.1 & 222.9& 4.91 & 0.0031 & 2.5 & 12.6 & 0.53 & 39.3 &495.1 \\
\end{tabular}
\end{ruledtabular}
\label{default}
\end{table*}%
\section{Conclusion}
\label{sec:CON}
We have established bulk superconductivity in a single crystal of Y$_3$Ru$_4$Ge$_{13}$ using magnetization, transport and heat capacity studies. Hall effect
studies show that the carrier concentration is 2 to 3 orders of magnitude smaller than that of typical metals like copper but comparable to that of
semimetal like arsenic. Moreover, we also observe an unusual temperature dependence of R$_H$ unlike the ones observed in High-T$_c$ cuprates and
the layered compound MgB$_2$. Preliminary band structure calculations\cite{r17} suggest Y$_3$Ru$_4$Ge$_{13}$ is a semimetal with a psuedo-gap in proximity 
of the Fermi level (i.e., a minimum in the density of states, with complex Fermi surfaces). Photoemission studies on a single crystal of  
Y$_3$Ru$_4$Ge$_{13}$ will help us to unravel the Fermi surface of this compound. In order to understand the usefulness of this material
for thermoelectric applications, thermopower and thermal conductivity measurements are in progress. Moreover, studies at ultra low temperature with high magnetic field
are planned to observe the elusive triplet superconductivity \cite{r13} in this crystal.

\end{document}